# A Neural Network-Based Scale-Adaptive Cloud-Fraction Scheme for GCMs

Guoxing Chen[1,2,3] , Wei-Chyung Wang[4] , Shixi Yang[1], Yixin Wang[1], Feng Zhang[1,2,3] , and Kun Wu[5]

[1]Department of Atmospheric and Oceanic Sciences & Institute of Atmospheric Sciences, Fudan University, Shanghai, China, [2]Shanghai Qi Zhi Institute, Shanghai, China, [3]Shanghai Frontier Science Center of Atmosphere-Ocean Interaction, Fudan University, Shanghai, China, [4]Atmospheric Sciences Research Center, University at Albany, State University of New York, Albany, NY, USA, [5]Key Laboratory of Meteorological Disaster, Ministry of Education, Collaborative Innovation Center on Forecast and Evaluation of Meteorological Disasters, Nanjing University of Information Science and Technology, Nanjing, China

**Abstract** Cloud fraction (CF) significantly affects the short- and long-wave radiation. Its realistic representation in general circulation models (GCMs) still poses great challenges in modeling the atmosphere. Here, we present a neural network-based (NN-based) diagnostic scheme that uses the grid-mean temperature, pressure, liquid and ice water mixing ratios, and relative humidity to simulate the sub-grid CF. The scheme, trained using CloudSat data with explicit consideration of grid sizes, realistically simulates the observed CF with a correlation coefficient >0.9 for liquid-, mixed-, and ice-phase clouds. The scheme also captures the observed non-monotonic relationship between CF and relative humidity and is computationally efficient, and robust for GCMs with a variety of horizontal and vertical resolutions. For illustrative purposes, we conducted comparative analyses of the 2006–2019 climatological-mean cloud fractions among CloudSat, and simulations from the NN-based scheme and the Xu-Randall scheme (optimized the same way as the NN-based scheme). The NN-based scheme improves not only the spatial distribution of the total CF but also the cloud vertical structure. For example, the biases of too-many high-level clouds over the tropics and too-many low-level clouds over regions around 60°S and 60°N in the Xu-Randall scheme are significantly reduced. These improvements are also found to be insensitive to the spatio-temporal variability of large-scale meteorology conditions, implying that the scheme can be used in different climate regimes.

**Plain Language Summary** Cloud fraction (CF) is crucial to cloud climate effects of both reflecting shortwave radiation and absorbing/emitting longwave radiation. However, the simulation of CF in general circulation models (GCMs) has been difficult, because most clouds are smaller than the typical scales of GCM grids and cannot be resolved by the grid-scale physics, while the physical understanding of sub-grid processes is still inadequate. Thus, this study uses a data-driven approach, that is, a neural network, to parameterize the sub-grid CF in climate models. The database for training and evaluating this NN-based scheme is obtained by upscaling the CloudSat (quasi-) observational data and emulating the GCM "grid-mean" properties required for cloud-fraction parameterization, minimizing the data-oriented biases. Moreover, the effects of the GCM horizontal and vertical grid sizes are both considered in the network, increasing the scheme adaptivity for use in GCMs with different resolutions. Results show that the NN-based scheme correctly predicts observed features of cloud-fraction variation with cloud condensate content and relative humidity for clouds of different phases and better predicts total cloud-fraction spatial distribution and cloud vertical structure than the conventional Xu-Randall scheme. This suggests that the NN-based scheme has the potential to reduce the biases of cloud radiative effects existing in current GCMs.

## 1. Introduction

Clouds play important roles in the Earth climate system. They dominate the energy budget by reflecting shortwave radiation and trapping longwave radiation (Wild et al., 2019), participate in the hydrological cycle via precipitation, and alter mass and energy vertical profiles by cloud venting (G. Chen et al., 2012; Yin et al., 2005) and latent-heat release. On the other hand, clouds are strongly coupled with aerosols and meteorology (Stevens & Feingold, 2009), involving complex feedbacks that span several temporal and spatial scales (e.g., G. Chen, Wang, et al., 2018; G. Chen, Yang, et al., 2018; Lau et al., 2006; Song et al., 2019; Xue et al., 2008). However, clouds are sub-grid scale in nature for current climate models, and cloud macro- and micro-physical properties in models have to be parameterized using the grid-mean atmospheric properties, making clouds the main source of most uncertainties in studies on climate change and climate modeling (e.g., Caldwell et al., 2016; Stevens et al., 2016).

Specifically, cloud-fraction parameterization poses a major challenge in climate modeling. Findings in the recent Coupled Model Intercomparison Project Phase 6 (CMIP6) revealed that current general circulation models (GCMs) have biases in both the daily-mean CF and the cloud diurnal variation. For example, for the former, the simulated CF is too small over the tropics, extra-tropics, and midlatitude regions (Li et al., 2021; Vignesh et al., 2020); and for the latter, the CF over land is too small during the daytime and too large during the nighttime (G. Chen et al., 2022; G. Chen & Wang, 2016a). As the shortwave cloud radiative effect (SWCRE) occurs only during the daytime while the longwave cloud radiative effect (LWCRE) persists throughout the daytime and nighttime, the biases in CF can affect not only the energy balance but also the energy diurnal variation, which may have implications on atmospheric variations of longer time scales (e.g., Ruppert, 2016; Slingo et al., 2003).

Current parameterization schemes of CF can be divided into diagnostic and prognostic approaches. In the diagnostic approach, the sub-grid CF is usually a function of the instantaneous grid-mean properties such as atmospheric stability, relative humidity, and/or cloud water content (e.g., Shiu et al., 2021; Sundqvist et al., 1989; Xu & Randall, 1996). These schemes are easy to implement and computationally efficient but may be too simple. On the other hand, the prognostic approach, which explicitly simulates the temporal variance of CF using source and sink terms associated with advection, cumulus convection, stratiform condensation, evaporation, and precipitation (Park et al., 2016; Tiedtke, 1993; Tompkins, 2002; Wilson et al., 2008), seems more physical. However, all source/sink terms are empirically related to the grid-mean properties, which are difficult to verify with observations. In addition, climate models usually employ different horizontal and vertical resolutions, which may affect the sub-grid statistical characteristics and thus the cloud-fraction parameterization.

In recent years, deep-learning methods have been demonstrated to be an effective alternative approach in atmospheric modeling (Chantry et al., 2021; Schultz et al., 2021). They can help reveal associations among atmospheric parameters by learning directly from data, regardless of inadequate or even none prior domain knowledge. For example, the feedforward neural network (also called multiple-layer perceptron, hereafter denoted as neural network for short) is good at emulating complex functions and can be used to replace certain modules by either accounting for poorly-understood processes or saving computational cost. It has been used to parameterize processes such as cloud cover (Grundner et al., 2022), radiation (Krasnopolsky & Fox-Rabinovitz, 2006), convection (Han et al., 2020; Rasp et al., 2018; T. Zhang et al., 2021), and boundary-layer turbulence (J. Wang et al., 2019). All yield promising outcomes. For the same reasons, the networks with complex architectures exhibit superiorities in building tools for specific predictions such as El Niño (Ham et al., 2019; Nooteboom et al., 2018), precipitation (G. Chen & Wang, 2022; Ravuri et al., 2021; Shi et al., 2015) and clouds (J. Zhang et al., 2018), and for data processing (Kim et al., 2021; Leinonen et al., 2021; Pan et al., 2019, 2021, 2022; Rasp & Lerch, 2018; Y. Zhang et al., 2021).

In this study, we introduce a neural network-based (NN-based) diagnostic scheme for simulating CF in climate models. This scheme has advantages in at least three aspects. First, using the neural network avoids the non-physical assumption of closed-form expressions in the parameterization; second, the data for developing the scheme were obtained from CloudSat observations of cloud profiles and the associated large-scale meteorology conditions, minimizing the data-oriented scheme biases; and third, the scale adaptivity (both horizontally and vertically) is considered while developing the scheme so that the scheme has higher robustness across models with different resolutions and in models with varying resolutions such as the Model for Prediction Across Scales-Atmosphere (MPAS; Skamarock et al., 2012) and the Global-to-Regional Integrated forecast SysTem atmosphere model (GRIST; Y. Zhou et al., 2020).

The rest of this manuscript is arranged as follows: Section 2 describes the data preparation; Section 3 introduces the NN-based scheme and the conventional Xu-Randall scheme (Xu & Randall, 1996), the latter of which is used as a baseline for evaluating the NN-based scheme; Section 4 tests the NN-based scheme in aspects of accuracy and scale adaptivity within the context of a machine-learning study while Section 5 evaluates the NN-based scheme with multiple-year CloudSat data using an offline application; and lastly, the summary and discussion are given in Section 6.

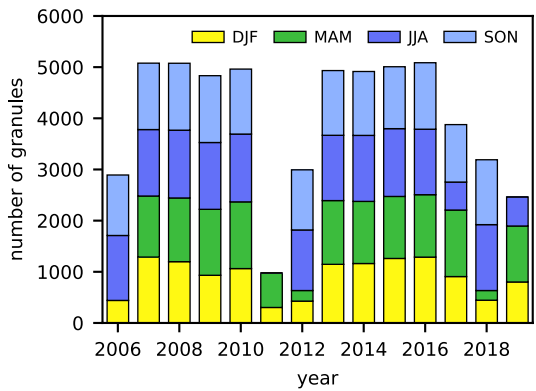

**Figure 1.** Year-by-year variation of CloudSat data amount used in this study. The colored segments indicate the portions of winter (December–January–February), spring (March–April–May), summer (June–July–August), and autumn (September–October–November), respectively.

## 2. Data Preparation

### 2.1. CloudSat

We use the CloudSat data (available at https://www.cloudsat.cira.colostate.edu) for training, validation, and testing of the NN-based cloud-fraction scheme. CloudSat is a polar-orbiting satellite launched by NASA in April 2006. It observes cloud vertical structure along its track using a 95 GHz radar onboard (Stephens et al., 2002). Its data have been widely used previously in studies on cloud climatology (e.g., T. Chen et al., 2016; Kato et al., 2010; Tang et al., 2020; Weisz et al., 2007), model development (e.g., Di Giuseppe & Tompkins, 2015; Li et al., 2018), and model evaluations (e.g., Bodas-Salcedo et al., 2008; Greenwald et al., 2010; Kodama et al., 2012; Vignesh et al., 2020; M. Wang & Zhang, 2018).

This study uses six variables from three CloudSat products: cloud liquid and ice water contents from 2B-CWC-RO (Austin & Wood, 2018); cloud profiling radar (CPR) cloud mask from 2B-GEOPROF (Marchand et al., 2008); atmospheric pressure, temperature, and specific humidity from ECMWF-AUX (Partain, 2022). The data all have a horizontal resolution of 1.1 km and a vertical resolution of 240 m.

Figure 1 presents the availability of CloudSat data at the time of our analysis. As CloudSat completes around 14.6 revolutions around Earth per day, there are around 5,000 granules of observations annually in normal years like 2007–2010 and 2013–2016. In 2011–2012 and 2017–2019, the CloudSat observation was interrupted occasionally due to reasons such as battery anomalies or orbit maneuvers (https://cloudsat.atmos.colostate.edu/news/CloudSat_status), yielding fewer data in these years. Meanwhile, the granule amount is relatively larger in the boreal summer and autumn than in the winter and spring. This may imply that the data is less representative of cloud climatology in the winter and spring seasons, but it does not affect the robustness of our scheme. It is shown below that the available data are more than enough for the scheme development and evaluation.

### 2.2. Data Preprocessing

Typical GCM grids have horizontal sizes ($\Delta x$) of 10–10s km (much larger than the CloudSat horizontal resolution) and vertical sizes ($\Delta z$) of 0.1–1s km (mostly larger than the CloudSat vertical resolution). Thus, the CloudSat clouds can be considered as sub-grid clouds at GCM grids, and the CloudSat data can be upscaled/aggregated to emulate atmospheric properties and CF simulated at GCM grids for the input and output of a cloud-fraction parameterization scheme.

During upscaling, the "grid-mean" temperature, specific humidity, atmospheric pressure, and liquid/ice water contents are calculated by averaging the raw CloudSat data within the range of $\Delta x \times \Delta z$, while the sub-grid CF is calculated as the horizontal cloud area fraction (i.e., each profile in the cloudy subarea must have one or more of the 240-m layers with CPR cloud mask ≥30 within $\Delta x \times \Delta z$). The definition of this sub-grid CF is consistent with that in most radiation parameterizations, and the resulting values are usually larger than the cloud volume fraction required by certain microphysical parameterizations (Grundner et al., 2022). More details of the upscaling process can be found in Y. Wang et al. (2023). In addition, to account for the different horizontal and vertical resolutions of GCMs, the upscaling is carried out for 42 different $\Delta x - \Delta z$ combinations. Therein, $\Delta x$ spans 40–100 km with an increment of 10 km while $\Delta z$ spans 10–60 hPa with an increment of 10 hPa, covering most grid sizes that may be seen in current GCMs.

It is noticed that the amount of available CloudSat data is far more than enough for the scheme development. For example, upscaling the data in 2015 to the resolution of 100 km × 20 hPa (hereafter x100z20 for short) can yield more than 10 million cloudy samples, which is more than sufficient for training a neural network with a simple architecture and relatively-few trainable parameters (see Section 3.1). Therefore, to save training time, we randomly draw 100 granules from the data in 2015 when upscaling to each set of resolutions. The choice of 2015 is arbitrary and using the data in any other year or multiple years does not affect the training results. Upscaling to 42 sets of resolutions yields more than 10 million cloudy samples in total. These samples are randomly split into three parts: 60% for scheme training, 20% for scheme validation, and 20% for scheme testing.

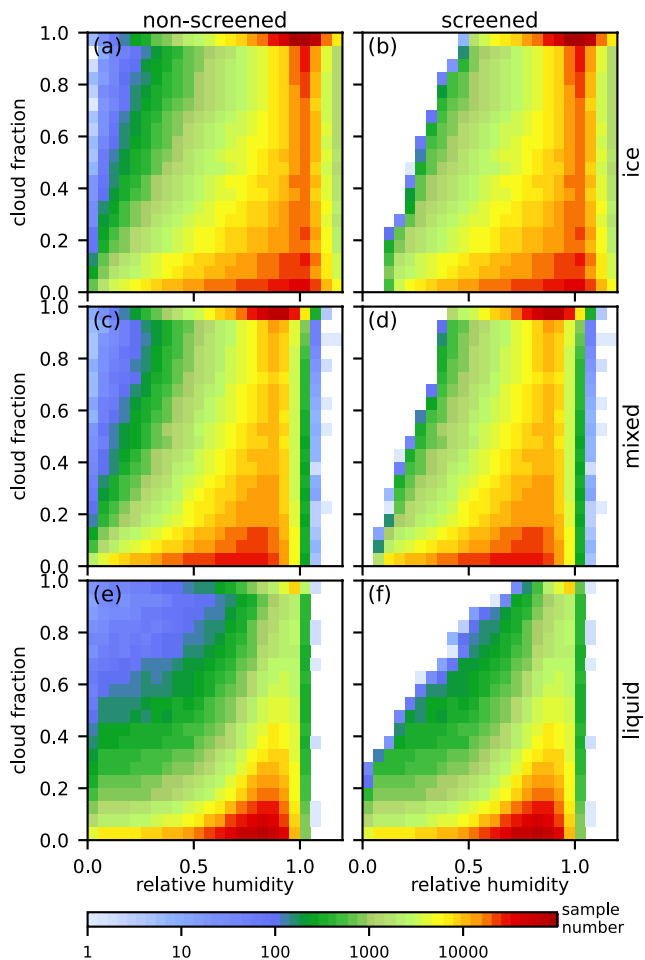

**Figure 2.** Effect of the isolation-forest screening on the data population. The shadings indicate sample numbers in the upscaled CloudSat data at the resolution of x100z20 for the whole year of 2015. An isolation-forest model is trained for clouds of each phase (i.e., ice, mixed, and liquid). The inputs of the models are the relative humidity of the respective phases and the sub-grid cloud fraction, and the contamination ratio is set to 0.01.

### 2.3. Data Uncertainty

The CloudSat data uncertainty exists in two aspects. First, the uncertainty of satellite data retrieval has been discussed in previous studies (e.g., Kotarba & Solecki, 2021; Weisz et al., 2007). Notably, the radar-only product of 2B-GEOPROF (2B-CWC-RO) is less reliable for cloud mask (cloud water contents) than the radar-lidar (radar-visible) combined product of 2B-GEOPROF-Lidar (2B-CWC-RO), respectively. Here, we intentionally avoid using the latter two products to ensure the data inherent consistency to the most degree. The radar-lidar combined cloud mask is not consistent with cloud water contents in 2B-CWC-RO while the radar-visible combined water contents are only available for the daytime and thus less representative.

Second, the uncertainty of atmospheric states from the ECMWF-AUX is remarkable but seldom discussed. Particularly, it was revealed that the upscaled CloudSat data have some anomalous samples with large cloud cover at very-low relative humidity (Y. Wang et al., 2023), which is counterintuitive and unrealistic. These are attributed to biases in the ECMWF-AUX data. For example, the ECMWF model tends to underestimate the inversion layer height in the eastern subtropical ocean (e.g., the southeast Pacific; Andrejczuk et al., 2012; G. Chen et al., 2015; G. Chen & Wang, 2016b), causing underestimated relative humidity in the upper cloud layer. As there is not a well-defined quantitative relationship between the sub-grid CF and the grid-mean relative humidity, it is difficult to find a physical-based method to detect these anomalous data.

Therefore, to facilitate the process of detecting anomalous samples in the upscaled data, we used an off-the-shelf machine-learning method—isolation forest (Liu et al., 2008, 2012). The method assumes that anomalous samples are isolately distributed from normal samples in the sample space and thus easier to be isolated than normal samples through random partitioning of the data. The algorithm builds a collection of binary trees from random subsets of data, where the anomalous samples should lead to shorter paths to leaves than normal samples, and aggregates the anomaly score from each tree to come up with a final anomaly score for determining the anomalous samples.

In this study, we train an isolation-forest model for clouds of each phase (i.e., ice, mixed-phase, and liquid) using 100,000 samples randomly drawn from the CloudSat data upscaled at the resolution of x100z20 for the whole year of 2015. The inputs for the models are sub-grid CF and relative humidity for the respective phases, where the saturated water vapor mixing ratio for the mixed phase is set to the weighted mean of those over ice and liquid surfaces. The contamination ratio, a parameter setting the proportion of anomalous samples in the dataset, is set to 0.01 for clouds of all three phases.

Figure 2 compares the sample number distributions of the non-screened and screened datasets. The anomalous samples exist in clouds of all three phases (Figures 2a, 2c, and 2e). The screening only removes samples with large CF but small relative humidity and does not affect the rest samples (Figures 2b, 2d, and 2f).

To further demonstrate the effect of screening, we examine an example of the upscaled CloudSat observation by the granule No. 03307. Shown in Figure 3, many samples in the upper portion of the cloud layer have the sub-grid CF (numbers in the figure) larger than 0.5 and the relative humidity (colored shadings in the figure) lower than 0.2, and the screening successfully distinguishes these anomalous samples (marked with red strikethroughs).

Note that the data screening is only to remove samples that we consider to be unrealistic. The data screening certainly does not affect the resulting scheme simply because the anomalous samples take a very small fraction of the CloudSat data (Figure 2). Likewise, increasing the contamination ratio also has little effect on the scheme results over the normal samples that dominate the data population. For example, setting the contamination ratio to 0.05 further reduces the number of samples having large CF at small relative humidity but does not cause any evident changes to the variation of CF with either relative humidity or cloud condensate mixing ratio (shown in Figures S1–S2 in Supporting Information S1).

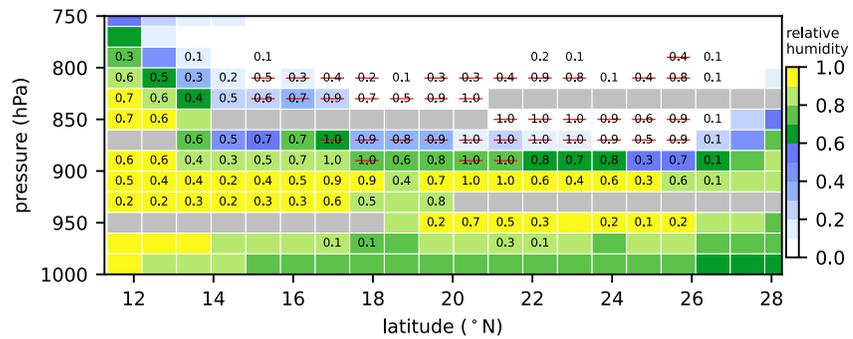

**Figure 3.** Effect of the data screening on data samples, demonstrated using the CloudSat observation at 12°–28°N by the granule No. 03307 on 11 December 2018, upscaled at x100z20. Numbers in the figure indicate observed sub-grid cloud fraction, colored shadings indicate relative humidity over the liquid surface, and gray shadings indicate no valid observations. The detected anomalous samples are marked with red strikethroughs.

## 3. Scheme Description

This section describes the neural network architecture and the algorithm of the Xu-Randall scheme. The database is the upscaled CloudSat data for 42 sets of resolutions (Section 2.2). The training, validation, and test datasets include around 5.4, 1.8, and 1.8 million samples, respectively. The former two datasets are used for the network-based scale-adaptive (NSA) scheme development and the Xu-Randall scheme tuning, while the test dataset is used for the scheme testing in Section 4.

### 3.1. Architecture of the Neural Network-Based Scale-Adaptive Cloud-Fraction Scheme

Figure 4 presents the architecture of the NN-based scale-adaptive (NSA for short) cloud-fraction scheme (Figure 4a). The network is composed of one input layer, a tunable number $n$ of hidden layers, one output layer, and the ReLU activation layers (not shown in the figure) following the input layer and hidden layers. The neuron amounts in each hidden layer ($N_{hi}, i = 1, 2, …, n$) are also tunable hyperparameters. The loss function is the mean square error.

The input layer consists of eight variables: air pressure ($P$), air temperature ($T$), liquid/ice water mixing ratios ($Q_l/Q_i$), relative humidity over liquid/ice surfaces ($R_l/R_i$, calculated with the upscaled atmospheric pressure, temperature, and specific humidity), and the horizontal/vertical grid sizes ($\Delta x/\Delta z$) of the host GCM. Therein, the first six variables are chosen because of their close association with cloud formation based on domain knowledge. They are also used in the Xu-Randall scheme, making it fair to compare results from the NSA and the Xu-Randall schemes in the following sections. $\Delta x$ and $\Delta z$ are to implicitly include the effects of grid sizes on the sub-grid cloud 3D structure (e.g., cloud sizes and vertical overlap) and thus the sub-grid cloud cover. As shown below, it provides the scheme with the flexibility of use at a variety of GCM resolutions.

We are aware that the input size could be reduced by replacing $R_l$ and $R_i$ using the specific humidity ($Q_v$). However, using $Q_v$ causes a larger burden for the neural network, and the resulting network should have more hidden layers or more neurons to get training accuracy close to that of the network using $R_L$ and $R_I$. Thus, using $R_L$ and $R_I$ is a choice out of computation efficiency. The output layer has only one neuron, that is, the sub-grid cloud fraction (CF).

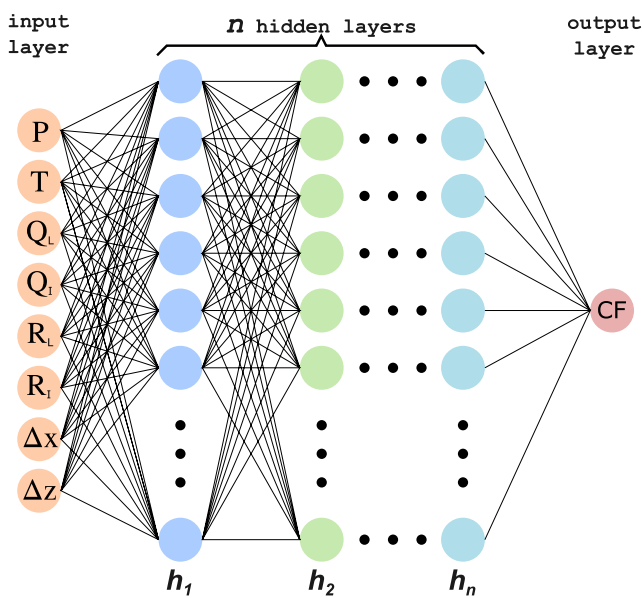

**Figure 4.** Architecture of the neural network-based scale-adaptive cloud-fraction parameterization. The output of the network is bounded to 0–1 before the analysis.

We utilize one single network to parameterize the CF of different phases for simplicity. As the CF variation with $Q_C$ and relative humidity is sensitive to the cloud phase (Y. Wang et al., 2023; also shown below in Figure 6), using separate networks for clouds of different phases may yield better results. However, using multiple networks increases the coding difficulty and the memory cost for storing the network parameters. Thus, the multiple-network approach is not taken in this study.

The hyperparameters are determined by minimizing the validation accuracy. We have tested the neuron number in each hidden layer (8, 16, 32, 64, 128, 256, 512), the optimizer (Adam vs. SGD), the learning rate (0.1, 0.01, 0.001), the batch size (256, 512, 1,024) and the activation function (Sigmoid vs. ReLU) when preparing this manuscript. In the end, we set $n$ to 2, $N_{h1}$ and $N_{h2}$ both to 128. The network has 17,793 trainable parameters and yields an root-mean square error (RMSE) of 0.130 over the validation dataset. Below, the scheme testing and the offline application of the NSA scheme are both based on this setting.

### 3.2. Tuned Xu-Randall Scheme

The Xu-Randall scheme is a conventional diagnostic cloud-fraction parameterization scheme for use in climate models. It is the default cloud-fraction scheme in the Weather Research and Forecast (WRF) Model and has been used in many weather and climate studies (e.g., G. Chen et al., 2021; G. Chen, Yang, et al., 2018; Song et al., 2019; Yang et al., 2018). The scheme is also employed in the FGOALS-f3 GCM developed by the Institute of Atmospheric Physics, Chinese Academy of Sciences (L. Zhou et al., 2015). We obtained its code from the WRF Model Version 3.7.1 and use it as a baseline for evaluating the NSA scheme. Below we briefly introduce the Xu-Randall scheme while more details can be found in the paper by Xu and Randall (1996).

The Xu-Randall scheme was built based on the simulations of a cloud ensemble model using a curve-fitting approach. It assumes the sub-grid CF is a function of the grid-mean relative humidity ($R$) and cloud condensate mixing ratio ($Q_C$, i.e., $Q_I$ for ice clouds, $Q_L$ for liquid clouds, and $Q_I + Q_L$ for mixed-phase clouds). The function formula is

$$\text{CF} = R^{\beta} \left[ 1 - \exp\left( - \frac{\alpha Q_C}{[(1-R)Q*]^{\gamma}} \right) \right], \text{if } R < 1 \qquad (1)$$

$$\text{CF} = 1, \text{if } R \geq 1 \qquad (2)$$

where $Q*$ is the grid-mean water vapor saturation mixing ratio, $\alpha = 100$, $\beta = 0.25$, and $\gamma = 0.49$ are empirical parameters determined through curve-fitting based on the simulation dataset.

As the Xu-Randall scheme was built on a completely different dataset than the NSA scheme, it would be unfair to expect the scheme produces results close to the CloudSat observations. Therefore, we have tuned the scheme using the same procedure and the same datasets (i.e., the aforementioned training and validation datasets) that are used for training the NSA scheme, yielding $\alpha = 70.3378$, $\beta = 0.0507$, and $\gamma = 0.6315$. Below we mainly present results from the tuned Xu-Randall scheme, and results of the original Xu-Randall scheme are also included in Supporting Information S1 for interested readers.

## 4. Scheme Testing

This section evaluates the NSA scheme using the test dataset within the context of a machine-learning study. Results of the cloud spatial and temporal characteristics predicted by the scheme are given in Section 5.

### 4.1. Accuracy

Figure 5 presents the joint probability density distribution of CFs from the scheme predictions and the CloudSat observations for clouds of different phases, estimated based on the test dataset. The NSA scheme is shown to well predict CFs for clouds of all three phases and has relatively larger biases (underestimation) when the observed

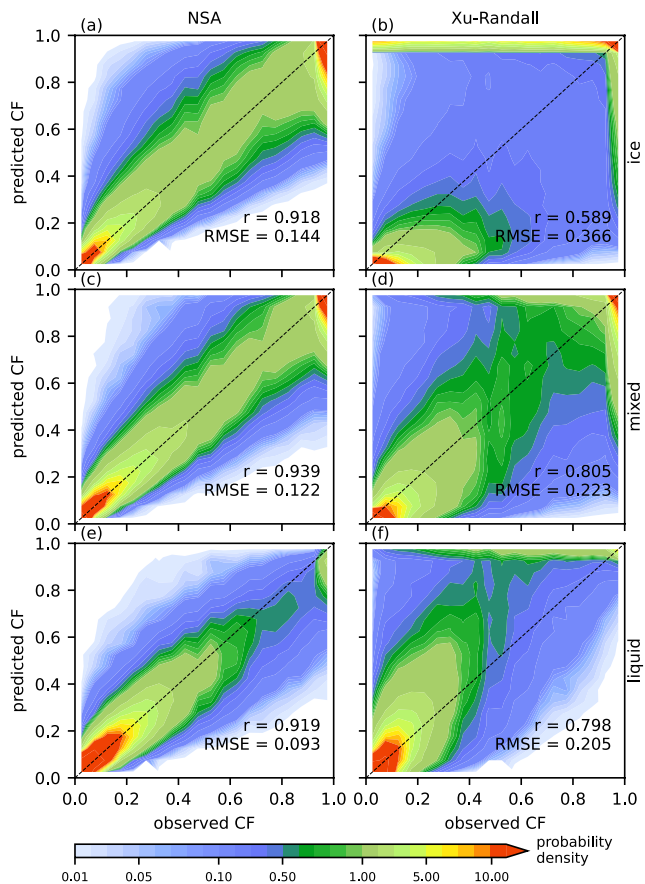

**Figure 5.** Joint probability density distribution between cloud fractions from the scheme predictions and the CloudSat observations for clouds of different phases (ice-only, mixed, and liquid-only), estimated based on the test dataset: (left) results from the network-based scale-adaptive (NSA) scheme; and (right) results from the tuned Xu-Randall scheme. The lower-right numbers indicate the correlation coefficient (r) and root-mean square error (RMSE) of the predictions with respect to the observations for each type of clouds.

CF is close to 1 (Figures 5a, 5c, and 5e). This could suggest the effect of the data imbalance. However, we found that balancing the CF frequency distribution by undersampling the training data could not much improve the results. This implies that the large-CF samples may have larger uncertainties or larger inherent inconsistencies, making the scheme difficult to learn.

It is noticed that the correlation coefficient ($r$, larger is better) is the largest for the mixed-phase clouds, while the RMSE (smaller is better) is the smallest for the liquid clouds. The two metrics are not consistent for two reasons. First, the fraction of large-CF samples is much larger in ice and mixed-phase clouds than in liquid clouds. Large-CF samples contribute more to the RMSE than small-CF samples, causing the relatively larger RMSE in ice and mixed-phase clouds. Second, the CloudSat data for liquid clouds might have large uncertainties. Liquid clouds are mostly sited in the lower troposphere, where the atmospheric properties are dominated by turbulence and difficult to simulate by GCMs. Therefore, air temperature and humidity from the ECMWF-AUX may have large uncertainties in the lower troposphere, causing the smaller $r$ for the CF prediction of liquid clouds.

In contrast, the tuned Xu-Randall scheme yields larger RMSE and smaller correlation coefficient for clouds of all three phases. The biases over ice clouds are the largest, where the RMSE is 0.366, more than twice that of the NSA scheme, and the correlation coefficient is less than 0.6. It is noticed that the original Xu-Randall scheme reaches better results over the liquid clouds but worse results over the mixed and ice clouds (shown in Figure S3 in Supporting Information S1) than the tuned one. This indicates the tuning process has opposite effects on the scheme performance over liquid versus mixed and ice clouds and suggests that the sub-grid statistics could be sensitive to the cloud phases. Moreover, even when we tune the scheme individually for each upscaled resolution, the scheme still exhibits obvious inferiorities to the NSA scheme at all resolutions (shown in Figure S4 in Supporting Information S1).

Figure 6 shows the variation of CF with relative humidity and cloud condensate mixing ratio from the CloudSat observations and the scheme predictions. In the observations, the ice CF is dominated by $R$ (Figure 6a), the liquid CF is dominated by $Q_C$ (Figure 6i), and the mixed CF is sensitive to both factors (Figure 6e). It is worth highlighting that holding $Q_C$ unchanged, the CF variation with $R$ is non-monotonic for clouds of all three phases. The liquid CF decreases with $R$ until $R$ equals around 0.8, and increases afterward. In contrast, the ice CF increases with $R$ until $R$ equals around 1, and decreases afterward. The mixed CF exhibits a complex pattern perhaps due to the phase partition of $Q_C$. It is also noticed that the supersaturation (i.e., $R > 1$), which seldom occurs in liquid clouds, is not infrequent in ice clouds.

The NSA scheme prediction presents patterns quite close to the observed ones. Particularly, the scheme well captures the non-monotonic feature of the CF variation with $R$. For samples that dominate the data population (i.e., samples below the black contours), the scheme results are almost identical to the observations. The scheme may underestimate CF when $Q_C$ is large (especially for ice and mixed-phase clouds, Figure 6a vs. Figure 6b, Figure 6e vs. Figure 6f). Nevertheless, the occurrence frequency of the associated samples (i.e., samples above the black contours in Figures 6d, 6h, and 6l) is very low.

The tuned Xu-Randall scheme prediction presents similar patterns for clouds of three phases, in line with the same formula (i.e., Equations 1 and 2) across different cloud phases. The patterns for liquid and mixed clouds (Figures 6g and 6k) are close to the CloudSat observations, consistent with the relatively larger correlation coefficients and smaller RMSEs in Figures 5d and 5f. However, the pattern for ice clouds (Figure 6c) is quite different from the observed one, showing the scheme underestimates CF for $Q_C < 50$ mg kg$^{-1}$ and overestimates CF for $R > 1$. Results for the original Xu-Randall scheme are similar to the tuned one (figure not shown). Neither the original nor the tuned Xu-Randall scheme can capture the non-monotonic variation of CF with $R$, which is not included in the parameterization formula.

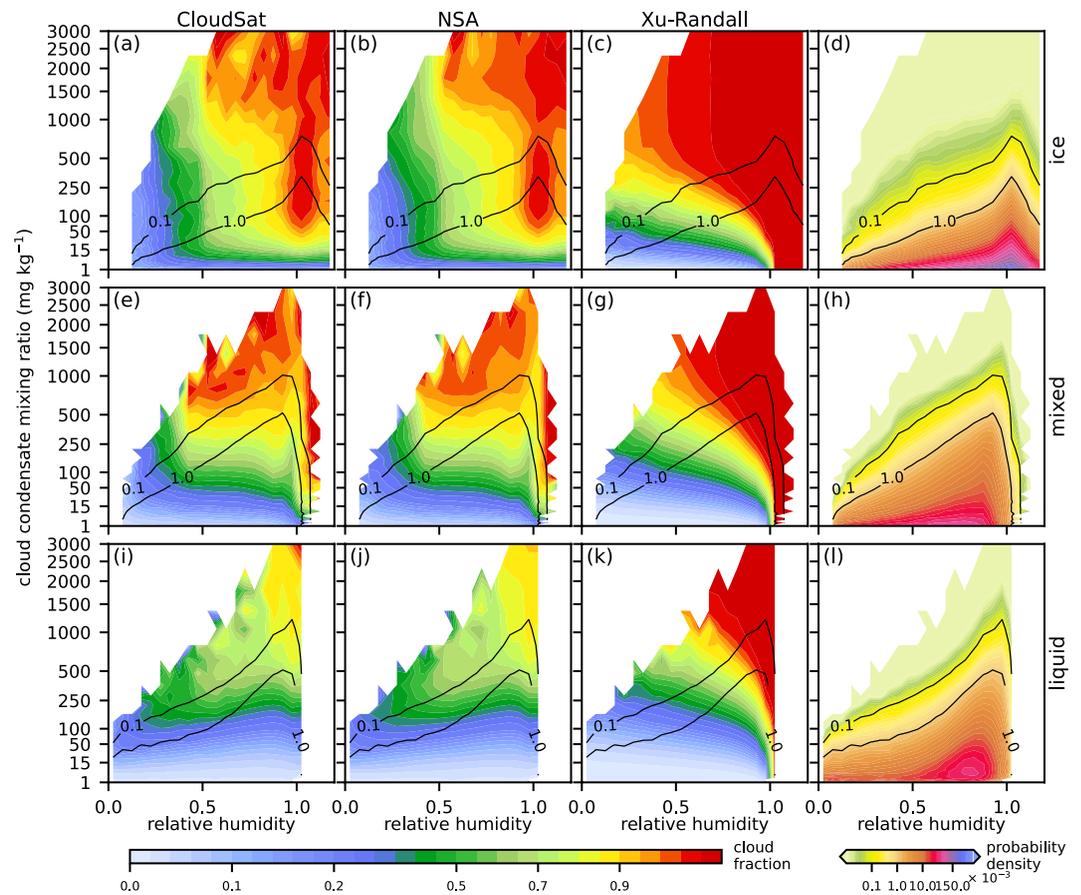

**Figure 6.** Comparisons of cloud fraction as a function of cloud condensate mixing ratio and relative humidity between the CloudSat observations (a, e, and i), the network-based scale-adaptive (NSA) scheme (b, f, and j), and the tuned Xu-Randall scheme (c, g, and k) for clouds of different phases (ice-only, mixed, and liquid only), estimated based on the test dataset. The right column (d, h, and l) indicates the joint probability density distribution of respective cloud types in the test dataset. Black lines indicate contours of the probability density of 0.001 and 0.0001, below which lie more than 88% and 97% of the corresponding sample populations, respectively.

## 4.2. Scale Adaptivity

To examine whether considering $\Delta x$ and $\Delta z$ improves the NSA scheme performance, we take two additional NN-based schemes as baselines: N and Nx100z20. Both schemes have the same network architecture as the NSA scheme except for excluding $\Delta x$ and $\Delta z$ from the input layer. The N scheme is trained based on the same dataset as the NSA scheme, while the Nx100z20 is trained based on the upscaled CloudSat at the resolution of x100z20 for the whole year of 2015 (no random drawing), where the sample amount is similar to that of the database for training the NSA scheme. Below we compare the three NN-based schemes based on the same test dataset that is used above in Figures 5 and 6.

Figure 7 compares the RMSE of the three NN-based schemes for samples with different horizontal and vertical resolutions. In the NSA scheme, the RMSE is not sensitive to $\Delta x$ or $\Delta z$ for ice clouds (Figure 7a), slightly sensitive to $\Delta z$ for mixed clouds (Figure 7d), and sensitive to both resolutions for liquid clouds (Figure 7g), where the RMSE tends to increase with $\Delta z$ and decreases with $\Delta x$. When $\Delta x$ and $\Delta z$ are excluded from the input layer, the N scheme has larger RMSEs for clouds of all three phases, especially the ice and mixed clouds (Figures 7b and 7e). As the deficiency of the N scheme may be attributed to the inherent inconsistency of the training database associated with different resolutions, we further examine the results of the Nx100z20, where the database inherent consistency is warranted. However, it shows that the RMSEs of the Nx100z20 scheme are even larger (Figures 7c, 7f, and 7i), and close to those of the NSA scheme only when $\Delta x$ is the largest and $\Delta z$ is the smallest. Therefore, we can conclude that both horizontal and vertical resolutions affect the sub-grid statistics of CF, and that including $\Delta x$ and $\Delta z$ in the scheme greatly increases the scale adaptivity of the NSA scheme, leading to higher robustness for use in GCMs with different horizontal and vertical resolutions.

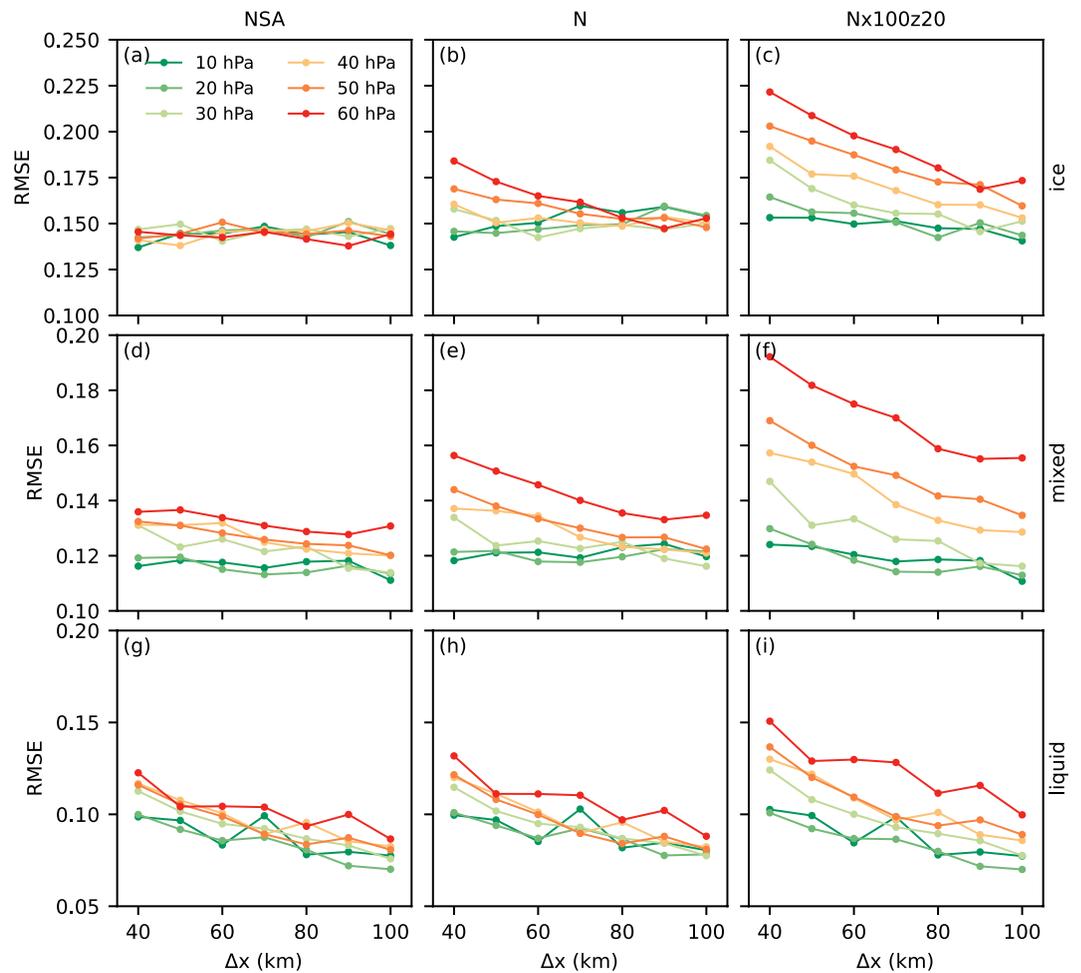

**Figure 7.** Comparisons of root-mean square error (RMSE) at different horizontal and vertical resolutions between the predictions of three neural network-based schemes estimated based on the test dataset. Both N and Nx100z20 schemes have the identical architecture to the network-based scale-adaptive (NSA) scheme except for excluding $\Delta x$ and $\Delta z$ from the input layer. The N scheme is trained with the same database as the NSA scheme, while the Nx100z20 scheme is trained based on the upscaled CloudSat data at the resolution of x100z20 for the whole year of 2015.

Figure 8 presents the mean CF at different horizontal and vertical resolutions from the CloudSat observations and the scheme predictions to show more details of how $\Delta x$ and $\Delta z$ may affect the CF parameterization. In the CloudSat observations (Figures 8a, 8e, and 8i), the CF tends to increase with $\Delta z$ and decrease with $\Delta x$. The NSA scheme well captures this feature, and the biases are less than 0.01 at all resolutions (Figures 8b, 8f, and 8j). In contrast, the biases of the N scheme are larger than 0.01 for most resolutions and can have values up to 0.09. Meanwhile, the biases are generally symmetric about the diagonal, that is, negative in the upper left and positive in the lower right. This is not a surprise. The scheme does not consider the resolution variability in the training database, and consequently the scheme biases tend to be the smallest in the middle of the resolution ranges and increase when the resolutions go toward either end. When we remove the inherent inconsistency associated with resolutions from the training database, the Nx100z20 scheme only shows better performance for samples with resolutions close to x100z20 (i.e., $\Delta x$ in 80–100 km and $\Delta z$ in 10–20 hPa). The bias increases rapidly with the decrease of $\Delta x$ and the increase of $\Delta z$, with the largest value of 0.14. Hence, it is further confirmed that the NSA scheme has good scale adaptivity.

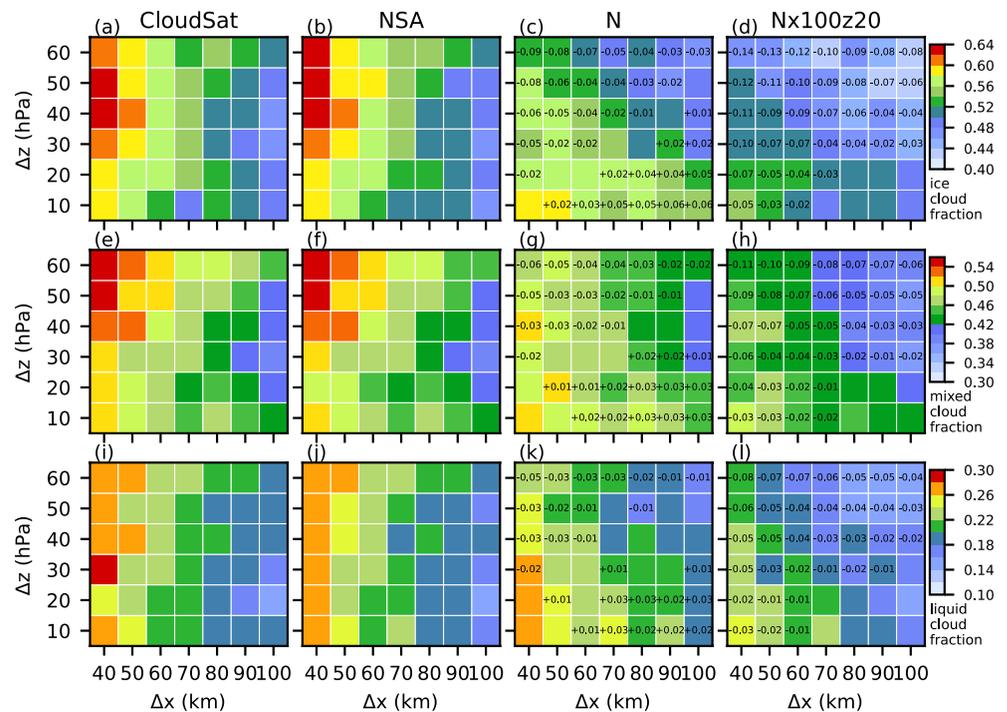

**Figure 8.** Comparisons of mean cloud fraction at different horizontal ($\Delta x$) and vertical ($\Delta z$) resolutions between the CloudSat observations and the predictions of three neural network-based schemes, estimated based on the test dataset. The numbers indicate the prediction-observation differences (prediction minus observation) with absolute values larger than 0.01.

## 5. Offline Application

This section compares the climatology of cloud spatial and temporal variability simulated by the NSA scheme with those from the CloudSat observations and the tuned Xu-Randall scheme using an offline method. The NSA and the tuned Xu-Randall schemes both take the upscaled CloudSat data at the resolution of x100z20 for the period of 2006–2019 as inputs to make the respective predictions. We first get an intuitive understanding of the scheme biases (Figure 9), then examine the generalizability of the NSA scheme (Figure 10), and last assess the scheme performance based on 2006–2019 multiple-year mean results (Figures 11–13).

We take the offline mode rather than incorporating the two schemes into a host model to exclude the distractions that could be caused by other model components (e.g., the simulated cloud condensate mixing ratios in the host model may deviate too much from the observations, Jiang et al., 2012) and possible feedbacks. Hence, the scheme-observation discrepancies can be mostly attributed to the scheme deficiencies, and the inter-scheme differences are not blurred by the possible compensating biases within the host model.

Figure 9 presents the observed and modeled cloud fractions in a CloudSat granule to get a better understanding of the data upscaling and an intuitive view of the scheme biases. In the raw CloudSat data (Figure 9a), a binary CF is determined by the CPR cloud mask: 1 for CPR cloud mask $\geq$30 and 0 otherwise. When upscaled (Figure 9b), the sub-grid CF has decimal values within 0–1. The upscaled CF may not have valid values at certain grids (indicated by light gray in Figure 9a) due to two reasons. First, the raw CloudSat does not have valid estimates for liquid or ice cloud water content; and second, none of the raw CloudSat data falls in the range of $\Delta z$ because the pressure change in 240 m exceeds $\Delta z$ (this situation occurs mainly in the lower atmosphere where the vertical gradient of atmospheric pressure is large).

The NSA and the tuned Xu-Randall schemes both well predict the general distribution of the sub-grid CF (Figures 9c and 9d) but exhibit different characteristics in the scheme-observation discrepancies. The NSA scheme tends to underestimate CF when CF is close to 1 (consistent with results in Figures 5 and 6), while the Xu-Randall scheme overestimates CF at many grids (particularly the high-level CF that are mostly ice clouds) because of its treatment of CF at supersaturation. The Xu-Randall scheme assumes that CF equals 1 when the grid-mean $R$ equals or exceeds 1 (Equation 2). This assumption works well for liquid clouds where the supersaturation is usually very low (Figure 6l), but not for ice clouds where the occurrence of large supersaturation is not scarce (Figure 6d).

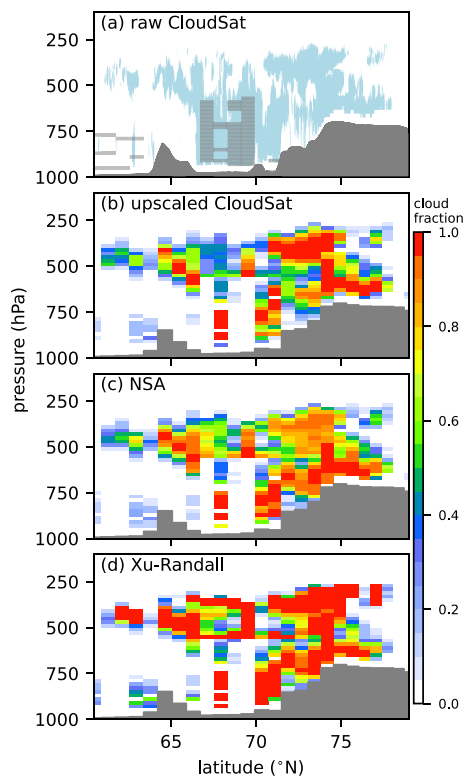

**Figure 9.** CloudSat observed cloud fraction (CF) at 60°–79°N by the granule No. 66364 on 13 October 2018 (a and b) compared with those from the network-based scale-adaptive (NSA) (c) and the tuned Xu-Randall (d) schemes. The dark gray regions at lower levels indicate the surface terrain. In (a), the light blue indicates overcast cloud coverage (cloud profiling radar cloud mask equals 30 or 40), in which the light-gray regions are grids where the CloudSat does not have a valid observation for one or more grid-mean properties that are required for parameterizing the sub-grid CF.

Figure 10 gives the year-by-year variation of the cloud-fraction RMSE, calculated with cloudy samples of all phases. Despite that the schemes are trained/tuned based on only data from the year of 2015, the year-by-year variations of both RMSEs are very small (both within the range of ±4% of the multiple-year mean) and do not show clear sensitivity to the interannual climate variability. Meanwhile, the RMSEs in 2011 (2006, 2012, and 2018), where the CloudSat data are dominated by granules in winter and spring (summer and autumn), present no unique features distinguished from results in other years. Therefore, it is inferred that the generalizability of the NSA scheme is very good and at least similar to that of the Xu-Randall scheme, warranting its robustness in different climates.

Figure 11 compares the 2006–2019 averaged global distributions of total CF from the observation and the scheme predictions, assuming the maximum-random overlap. The observed CF is larger in the tropics and the regions around 60° in both hemispheres and smaller in the subtropical regions. The two schemes both capture the general patterns. The tuned Xu-Randall prediction has a global mean of 0.52, larger than that from the original Xu-Randall scheme (0.46; Figure S5a in Supporting Information S1), which is closer to the observation (0.46) than the NSA prediction (0.43). However, the original/tuned Xu-Randall prediction has too large CF over the tropics (especially the India-Pacific warm pool) and too small/large CF over the Southern Ocean, where the NSA prediction is closer to the observation. Consequently, the NSA prediction has a larger spatial correlation coefficient (0.996) and a smaller RMSE (0.032) than the Xu-Randall predictions. Using the maximum overlap assumption reaches similar results while using the random overlap suggests that the tuned Xu-Randall scheme prediction is closer to the observation in the global-mean total CF (shown in Table S1 in Supporting Information S1). These conflicting results imply that the cloud vertical structure could be quite different between the scheme predictions, as examined below.

Figure 12 presents the zonal-mean vertical structure of CF from the observations and the scheme predictions. In the tuned Xu-Randall scheme, the CF overestimation in the tropics is associated with the too-large high-level CF (clouds above 440 hPa), while the CF overestimation in regions around 60°S and 60°N is due to the too-large low-level CF (clouds below 680 hPa). The original Xu-Randall scheme overestimates the high-level CF in the tropics as well but underestimates the low-level CF in regions around 60°S and 60°N (Figure S5b in Supporting Information S1). The tuning causes larger CF at all levels and yields better mid-level CF but worse low-level CF, which is consistent with the results on mixed and liquid clouds shown in Figure 5 and Figure S3 in Supporting Information S1. The NSA scheme predicts smaller high-level CF in the tropics and smaller mid- and low-level CFs in regions around 60°S and 60°N and reaches a larger spatial correlation (0.998) and a smaller RMSE (0.006) than both the original and tuned Xu-Randall schemes.

Figure 13 further examines the seasonal variations of the cloud vertical structure at different latitudes. In the observations, the cloud vertical structure varies with the seasonal migration at all latitudes. While both schemes can predict the general variation, the NSA scheme shows better results (i.e., larger correlation coefficient and smaller RMSE) than the original (Figure S6 in

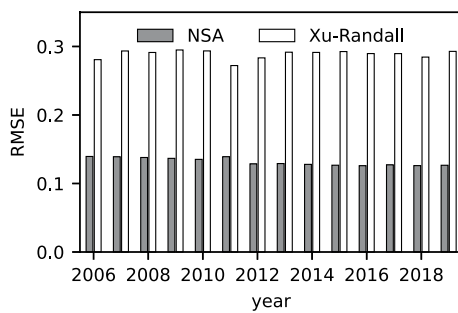

**Figure 10.** Cloud-fraction root-mean square error (RMSE) year-by-year variation of the network-based scale-adaptive (NSA) and the tuned Xu-Randall scheme predictions with respect to the CloudSat observations. Only cloudy grids are included in the calculation.

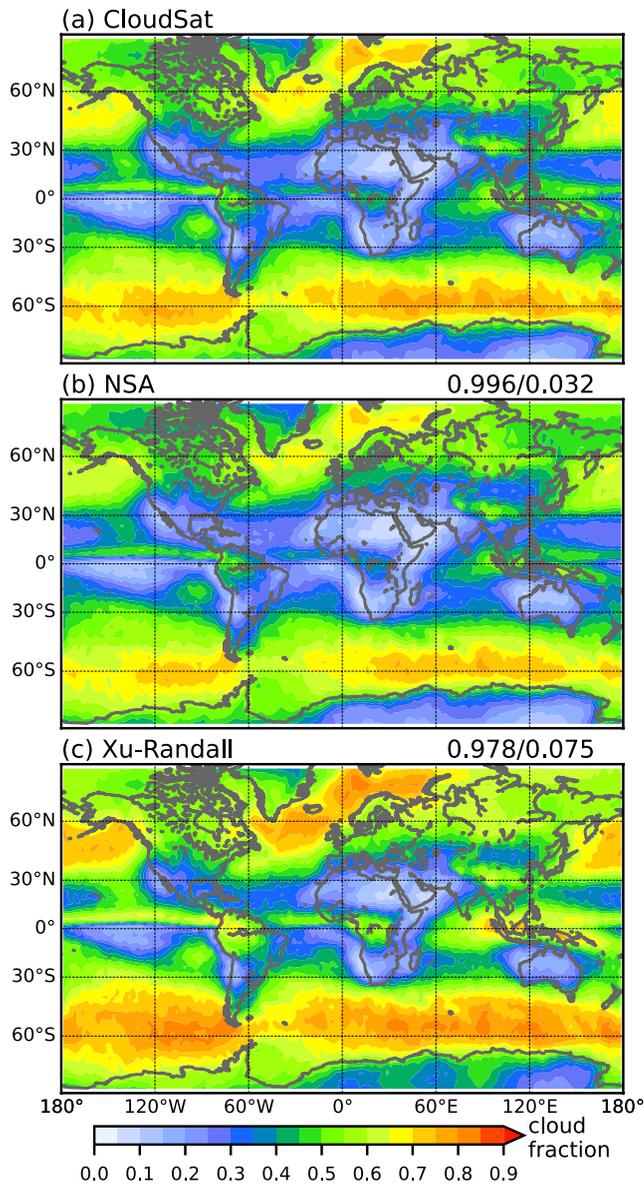

**Figure 11.** 2006–2019 mean global distribution of total cloud fraction (CF) (assuming maximum-random overlap) from the CloudSat observation, the network-based scale-adaptive (NSA) scheme, and the tuned Xu-Randall scheme. Numbers at the upper right corners of (b) and (c) indicate the correlation coefficient (*r*; before the slash) and root-mean square error (after the slash) of the respective scheme predictions with respect to the CloudSat observations. The global-mean total CF is 0.46, 0.43, and 0.52 for the CloudSat observations, the NSA, and the tuned Xu-Randall scheme predictions, respectively.

Supporting Information S1) and tuned Xu-Randall schemes in all seasons and all latitudes. This indicates that the NSA scheme generally exhibits superiorities in the spatial distribution of total CF and the cloud vertical structure, and further confirms that the superiorities are not sensitive to the spatio-temporal variability of the large-scale meteorology conditions, suggesting the scheme has enough robustness to simulate CF over different climate regimes.

## 6. Summary and Discussion

This study presents a neural network-based cloud-fraction parameterization scheme for use in climate models. The scheme is developed using the Cloud-Sat (quasi-)observational data, and for the first time considers explicitly the effects of both horizontal and vertical grid sizes on the cloud-fraction parameterization. It not only simulates realistically the cloud characteristics but also captures the observed non-monotonic functional relationship of CF with relative humidity. In addition, our preliminary study shows that using the NN-based scheme in the WRF Model hardly increases the computation time as compared with using the Xu-Randall scheme. The utility of the scheme in GCMs shows promising features in accuracy, scale adaptivity, and computational efficiency.

In the case of comparing with the Xu-Randall parameterization, the NN-based scheme simulates more-realistic total CF spatial distribution and cloud vertical structure. Particularly, the biases of too-large high-level CF over the tropics and too-small low-level CF over regions around 60°S and 60°N in the original Xu-Randall scheme, which agrees with the too-strong LWCRE and too-weak SWCRE over the respective regions as shown in current GCMs (G. Chen et al., 2022; Flato et al., 2013; Schuddeboom & McDonald, 2021), are much eased.

We are aware that the grid-mean atmospheric properties in GCMs could differ markedly from the CloudSat data (e.g., Jiang et al., 2012), hindering the above findings from being fully justified in GCMs. Hence, one of our undergoing studies is to further evaluate the NSA scheme using the ERA-Interim reanalysis (Dee et al., 2011), where the grid-mean relative humidity is assimilated with observations while the cloud condensates are fully model simulated. The preliminary results show that although cloud condensate mixing ratios in the reanalysis are much smaller than the CloudSat observations, using the NSA scheme still yields better mid- and high-level cloud fractions than the original Xu-Randall scheme (figure not shown). Therefore, it is believed that incorporating the NN-based scheme has the potential to improve the cloud radiative effects and the energy budget in GCMs.

One important aspect of CF is the vertical overlap, which has significant implications to cloud radiative effects (e.g., Liang & Wang, 1997; X. Wang et al., 2021; H. Zhang & Jing, 2016). As shown by F. Zhang et al. (2013), the inter-GCM spreads in cloud radiative effects can be largely attributed to the different treatments of cloud vertical overlap. In our scheme, this aspect, at least in the sub-grid scale, is implicitly considered, which reduces the simulation biases in CF (shown in Figures 7 and 8). Therefore, further investigation of the grid-scale cloud vertical overlap using similar approaches is plausible and warranted.

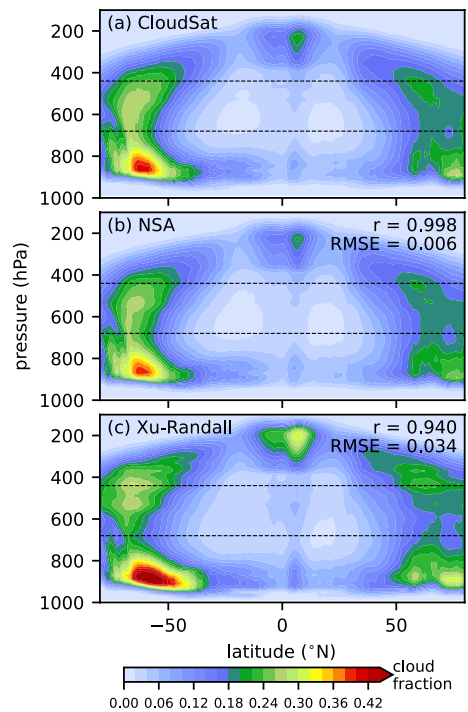

**Figure 12.** 2006–2019 averaged zonal-mean vertical distribution of cloud fraction from the CloudSat observations, the network-based scale-adaptive (NSA), and the tuned Xu-Randall scheme predictions. Numbers in (b) and (c) indicate the correlation coefficient (*r*) and root-mean square error of the respective scheme predictions with respect to the CloudSat observations. Dashed lines indicate the heights of 440 and 680 hPa, respectively.

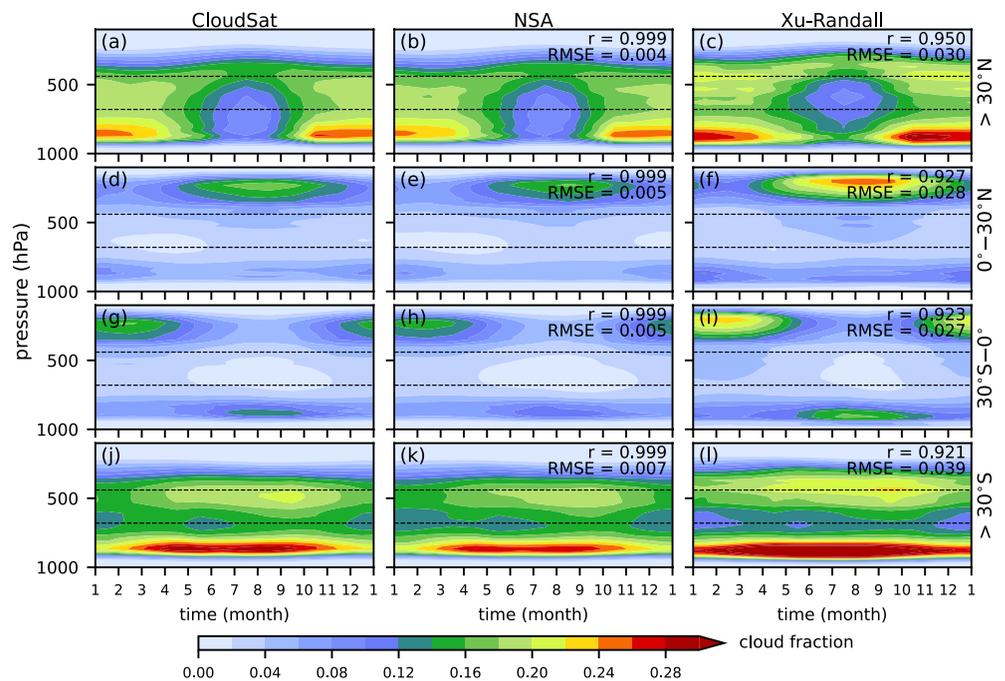

**Figure 13.** Annual variation of cloud vertical distribution averaged at different latitudes from the CloudSat observations (left), the network-based scale-adaptive (NSA) (middle), and the tuned Xu-Randall (right) scheme predictions. Numbers in the middle and right columns indicate the correlation coefficient (*r*) and root-mean square error (RMSE) of the respective scheme predictions with respect to the CloudSat observations. Dashed lines indicate the heights of 440 and 680 hPa, respectively.

## Appendix A: Calculation of Root–Mean Square Error and Correlation Coefficient

The root-mean square error (RMSE) and correlation coefficient ($r$) of a prediction with respect to the observation are calculated as follows:

$$\text{RMSE} = \sqrt{\sum_{i=1}^{M} (\text{pred}_i - \text{obs}_i)^2 / M} \tag{A1}$$

$$r = \frac{\sum_{i=1}^{M} \left( \text{pred}_i - \overline{\text{pred}} \right) \left( \text{obs}_i - \overline{\text{obs}} \right)}{\sqrt{\sum_{i=1}^{M} \left( \text{pred}_i - \overline{\text{pred}} \right)^2} \sqrt{\sum_{i=1}^{M} \left( \text{obs}_i - \overline{\text{obs}} \right)^2}} \tag{A2}$$

Therein, the pred and obs stand for the prediction and observation datasets, respectively, $i$ for the sample index, and $M$ for the sample amounts.

## Data Availability Statement

The neural network was implemented using the Pytorch application programming interface (Paszke et al., 2019). The codes for upscaling CloudSat data, scheme training, and result analysis together with the databases for training NSA and Nx100z20 schemes are preserved in G. Chen and Wang (2023).

**Acknowledgments**

This study is supported by the National Key R&D Program of China (2021YFC3000801) and the National Natural Science Foundation of China (42275074). WCW acknowledges the support from the SUNY Research Foundation Fixed-Price Balance Account (31972). The authors thank the editor (Dr. Tapio Schneider) and the two anonymous reviewers for their kind and patient comments, which greatly help clarify and improve this manuscript.